\documentclass[a4paper,12pt,useAMS,usenatbib]{mn2e}

\usepackage{graphicx}

\title[Miscentring in Galaxy Clusters]{Miscentring in Galaxy Clusters: Dark Matter to Brightest Cluster Galaxy Offsets in 10,000 SDSS Clusters}
\author[Zitrin et al.]{Adi Zitrin$^{1}$\thanks{E-mail:adizitrin@gmail.com}, Matthias Bartelmann$^{1}$, Keiichi Umetsu$^{2}$, Masamune Oguri$^{3}$, \and Tom Broadhurst$^{4,5}$\\\\\\
$^{1}$Institut f\"{u}r Theoretische Astrophysik, ZAH, Albert-Ueberle-Stra\ss e 2, 69120 Heidelberg, Germany\\
$^{2}$Institute of Astronomy and Astrophysics, Academia Sinica, P.~O. Box 23-141, Taipei 10617, Taiwan\\
$^{3}$Kavli Institute for the Physics and Mathematics of the Universe (Kavli IPMU, WPI), University of Tokyo, Kashiwa, Chiba 277-8583, Japan\\
$^{4}$Department of Theoretical Physics, University of the Basque Country, Vizcaya, 48940, Leioa, Spain\\
$^{5}$IKERBASQUE, Basque Foundation for Science, Alameda Urquijo 36-5, 48008 Bilbao, Spain}

\begin{document}


\pagerange{\pageref{firstpage}--\pageref{lastpage}} \pubyear{2012}

\maketitle

\label{firstpage}

\begin{abstract}

We characterise the typical offset between the Dark Matter (DM) projected centre and the Brightest Cluster Galaxy (BCG) in 10,000 SDSS clusters. To place constraints on the centre of DM, we use an automated strong-lensing (SL) analysis, mass-modelling technique which is based on the well-tested assumption that light traces mass. The cluster galaxies are modelled with a steep power-law, and the DM component is obtained by smoothing the galaxy distribution fitting a low-order 2D polynomial (via spline interpolation), while probing a whole range of polynomial degrees and galaxy power laws. We find that the offsets between the BCG and the peak of the smoothed light map representing the DM, $\Delta$, are distributed equally around zero with no preferred direction, and are well described by a log-normal distribution with $\langle \log_{10}(\Delta~[h^{-1} $Mpc$])\rangle=-1.895^{+0.003}_{-0.004}$, and $\sigma=0.501\pm0.004$ ($95\%$ confidence levels), or $\langle \log_{10}(\Delta~[\arcsec])\rangle=0.564\pm0.005$, and $\sigma=0.475\pm0.007$. Some of the offsets originate in prior misidentifications of the BCG or other bright cluster members by the cluster finding algorithm, whose level we make an additional effort to assess, finding that $\sim10\%$ of the clusters in the probed catalogue are likely to be misidentified, contributing to higher-end offsets in general agreement with previous studies. Our results constitute the first statistically-significant high-resolution distributions of DM-to-BCG offsets obtained in an observational analysis, and importantly show that there exists such a typical non-zero offset, in the probed catalogue. The offsets show a weak positive correlation with redshift, so that higher separations are generally found for higher-$z$ clusters in agreement with the hierarchical growth of structure, which in turn could potentially help characterise the merger, relaxation and evolution history of clusters, in future studies. In addition, the effective DM centre we adopt here, namely the peak of the smoothed light distribution representing the DM, can constitute a natural and alternative definition of cluster centers for optically-selected cluster catalogues.

\end{abstract}


\section{Introduction}\label{intro}

Galaxy clusters are the largest gravitationally-bound objects in the Universe, and are formed at later stages and relatively lower redshifts, in the hierarchical model. As such, galaxy clusters can shed light on the high end of the cosmic mass function and the evolution history of the Universe, and probe the acceptable cosmological model \citep[e.g.][]{Allen2011clusterparareview}.

Digitised large sky surveys such as the \emph{Sloan Digital Sky Survey} (SDSS; see \citealt{York2000SDSS_tech,Abazajian2003SDSS_I,Abazajian2009SDSS_7}), and increasing computational power, have driven in recent years statistical analyses of extensively large cluster samples. In these (mainly optical imaging) data, galaxy clusters are identified usually in an automated manner, via dedicated finding algorithms (e.g. \citealt{Postman1996Cat+finder,Kepner1999Finder,Gal2000Finder,GladdersYee2000Finder,Koester2007maxBCG_finder} and references therein; \citealt{Pierpaoli2011SDSScatalog}). In addition, recent developments allow an independent detection of unprecedentedly large numbers of galaxy clusters in the X-ray and Sunyaev-Zeldovich Effect (SZE) observations \citep[e.g.][]{Bohringer2004ClusterCat,Ebeling1998ClusterCat,SPT2010ClustersV1,SPT2011Clusters_V2,Planck2011ClustersV1}. Very large cluster samples have been used in many recent studies such as analyses of weak and strong lensing \citep[e.g.][]{Johnston2007WL_SDSS,Mandelbaum2008StackedWL,Zitrin2011d}, or to place constraints on the cosmological parameters \citep[e.g.][]{Rozo2010CosmoConstraintsSDSS}, and establish various scaling relations (e.g. \citealt{Rozo2009MassRichSDSS,Planck2011XSZ_scaling,Bauer2012massrichness}; see also \citealt{KravtsovBorgani2012Review}) as a few examples.

One of the major factors of noise or systematic uncertainty in these studies, especially when the clusters are identified optically and when the analysis is dependent on a predetermined centre, such as a stacked lensing analysis, cluster-background
galaxy cross-correlation measurement (\citealt{Johnston2007offset,Mandelbaum2010WL}; see also \citealt{Oguri2010_25clusters,Umetsu2011b}), is the ``misentring'' of the BCG with respect to the dark matter (DM). Such offsets may result from either a misidentification of the cluster finding algorithm, or by a real measured offset between the projected DM centre and the BCG \citep[see also][]{Johnston2007offset}, and will have a smoothing effect on the, e.g., stacked lensing signal (see \citealt{OguriTakada2011}). A unique way to trace the offsets of the correctly identified clusters (hereafter, for simplicity, we dub these \emph{true}, or \emph{real} offsets to distinguish them from the misidentified clusters), to high accuracy, is by a strong-lensing (SL) analysis, where usually the multiple images in the core are used to accurately map the galaxies and dark mass distributions \citep[e.g][]{Zitrin2009b,Umetsu2011}.

Due to the importance of the miscentring effect, there have been recent efforts to characterise its amplitude and size distribution. For example, \citet{Johnston2007offset} quantified this effect by using N-body simulation-based mock galaxy catalogues, and then running their cluster-finding algorithm (maxBCG in that case; \citealt{Koester2007maxBCG_finder}) to identify clusters in these catalogues, comparing the resulting BCG positions to the centres of the DM halos in the input simulations. They found that the offsets, i.e., for the clusters that are not centred on their BCG, are well described by a 2D gaussian of the form $P(R_{s})=\frac{R_{s}}{\sigma^{2}} \exp(-\frac{1}{2} (\frac{R_{s}}{\sigma})^{2})$, with $R_{s}$ being the magnitude of the offset and $\sigma=0.42~ h^{-1}$ Mpc, and that the effect they traced is indeed dominated by misidentified BCGs. More recently \citet{HilbertWhite2010} similarly found $\sigma=0.34-0.41~ h^{-1}$ Mpc (for different WMAP cosmologies). \citet{Johnston2007offset} also found that the fraction of misidentified clusters decreases with richness: $\sim60\%$ of the poorer ($N_{gal}\sim10$) clusters are correctly identified, versus $\sim85\%$ of the richer ($N_{gal}\sim100$) clusters. Although using these numerical simulations one can estimate the level and distribution of misidentified clusters, in such procedures the galaxies are assigned to DM halos in the simulation to begin with, so that the \emph{true} offset distribution, i.e. for clusters whose BCG was correctly identified, is hard to assess.

Observationally, several studies also examined the offsets between the BCG and the X-ray peak (or centroid), or recently, also with the SZ peak. For example, \citet{LinMohr2004} listed the BCG to X-ray peak offsets for a few dozen Two Micron All-Sky Survey (2MASS, \citealt{Jarrett2000_2mass}) clusters. They found, that about $75\%$ of the identified clusters lay within $0.06r_{200}$, and $90\%$ within $0.38r_{200}$, with a $\sim10\%$ contamination level of possibly misidentified BCGs. They suggested, that given these high fractions of small X-ray to BCG separations, the timescale for the BCG to sink to the cluster potential minimum may be short
compared to the relaxation timescale of the intracluster
gas. Or alternatively, the scale of merger required to perturb
the X-ray properties of the cluster could be smaller than
the scale of the merger required to offset the BCG from the cluster centre \citep{LinMohr2004}. \citet{MannEbeling2012} examined the X-ray peak and centroid offsets from the BCG, in 108 of the most luminous X-ray clusters, with the goal of constraining the evolution with redshift of
the cluster merger fraction, so that they also characterised the evolution of such offsets in redshift. They found that the distribution is roughly log-normal, and centred at 11.5 and 21.2 kpc for the offset of the BCG from the X-ray peak, and from the X-ray
centroid, respectively. In addition, they found an evolution for these offsets with redshift, so that higher separations are generally expected for higher redshift clusters, probably as a sign of higher merging fraction. \citet{Sehgal2012offsetsSZBCG} recently examined the SZ signal for 474 optically-selected (maxBCG) clusters and 52 X-ray selected (MCXC) clusters using data from the Atacama Cosmology Telescope (ACT). For the optically-selected sample, they found that the \emph{Planck} and ACT measurements could be explained if one assumed that the BCGs are offset from the SZ peaks uniformly between 0 and 1.5 Mpc. However, they point out that other factors (rather than the BCG offset) could be in play in explaining the observed discrepancies, especially since for the X-ray-selected sample, a much narrower BCG to X-ray peak offset distribution was found, peaking within 0.2 Mpc. In recent work, also, \citet{Song2012OffsetsSZ} find that the BCG to SZ centroid offset distribution, in 146 South Pole Telescope selected clusters with BCGs well identified in follow-up optical and near-infrared observations, is similar to that found previously in X-ray samples.

\citet{Oguri2010_25clusters} observationally examined the offset of the BCG from the centre of mass obtained in weak lensing (WL) analyses of a sample of 25 clusters. They found that the DM centre is overall consistent with that of the BCG (within $2\sigma$ level), and that the observed distribution can be described by two components. The first, significant component describing the small offsets, is a 2D Gaussian with $\sigma=0.09~ h^{-1}$ Mpc, and the second less-significant component describing the tail of larger separations, is fitted by the \citet{Johnston2007offset} finding: a 2D Gaussian with $\sigma=0.42~ h^{-1}$ Mpc. They concluded that about 10\% of their sample are BCG-offsetted clusters. \citet{Oguri2010_25clusters} also probed the same effect by complementary X-ray fits for their sample, but found no specific correlation as the X-ray centroids are usually more closely centred on the BCG (e.g. \citealt{LinMohr2004,Maughan2008clusterEv,MannEbeling2012}, although see also \citealt{Shan2010_38offsets}). They also note that the typical error on the mass centroid measurement in their WL analysis is $\sim50~ h^{-1}$ kpc, and there are a few clusters in their sample that have errors larger than $100~ h^{-1}$ kpc, which are non-negligible compared to the widths of the resulting 2D Gaussian distributions. In addition, \citet{Dietrich2012WLoffsets} recently found, by thoroughly simulating WL observations, that generally, the magnitude of peak-offsets in WL maps could primarily be a direct result of shape-measurement noise and smoothing of the WL mass maps. It is thus clear that for a more quantitative assessment of the distribution of BCG offsets, a much larger sample is needed, probed with sufficiently high resolution.

Recently, \citet{George2012offsets} examined in 129 galaxy groups (halo masses of up to $10^{14}~M_{\odot}$) in the COSMOS field, the offset between the WL halo centre and other tracers such as the BCG, most massive group galaxy, or X-ray centroid. They found that the BCG is one of the best tracers of the centre-of-mass, and is offset typically by less than 75 kpc from the (dark) halo centre. In addition, they concluded that similar offset distributions are highly susceptible to the centre definitions, i.e., centres defined following certain intensity centroids, can largely differ from those defined via the corresponding intensity peaks \citep[see][]{George2012offsets}, and that not accounting for the miscentring effect can cause a 5-30\% bias in stacked WL analyses. The X-ray to BCG offset magnitude is generally consistent with several previous studies mentioned therein \citep[e.g.][finding typically a few dozen kpc offsets or less]{Sheldon2001offsets,Koester2007maxBCG_cat,Sanderson2009locuss}.

\citet{Shan2010_38offsets} characterised the offsets between the X-ray peaks and lensing centres in 38 clusters (see also \citealt{Allen1998L-XrayDisc}). Although most clusters show small offsets, as is also usually seen in such lensing-analyses centre to BCG comparison \citep[e.g.][]{Smith2005,Richard2010locuss20}, about $45\%$ of their clusters, usually the merging, multiple-clump ones, show larger separations than $10\arcsec$, with a maximum of $\simeq54\arcsec$ (or $\sim200$ kpc, see also \citealt{Forero-Romero2010}). This, however, may be a result of either large fractions of unrelaxed clusters in their sample ($\sim60\%$), and, the ensemble of different SL techniques used for the comparison, many of which pre-assume or iterate for the DM centre while adopting a symmetric DM distribution (see references therein), which may be unrealistic given the perturbed and complex matter distribution seen especially in unrelaxed clusters. In that sense, such offsets or even the known discrepancy between mass estimates from lensing and X-ray \citep[e.g.][]{Allen1998L-XrayDisc,Richard2010locuss20}, may not be surprising \citep[see][]{Shan2010OriginOffsets}.

\begin{figure}
\centering
 \includegraphics[width=90mm]{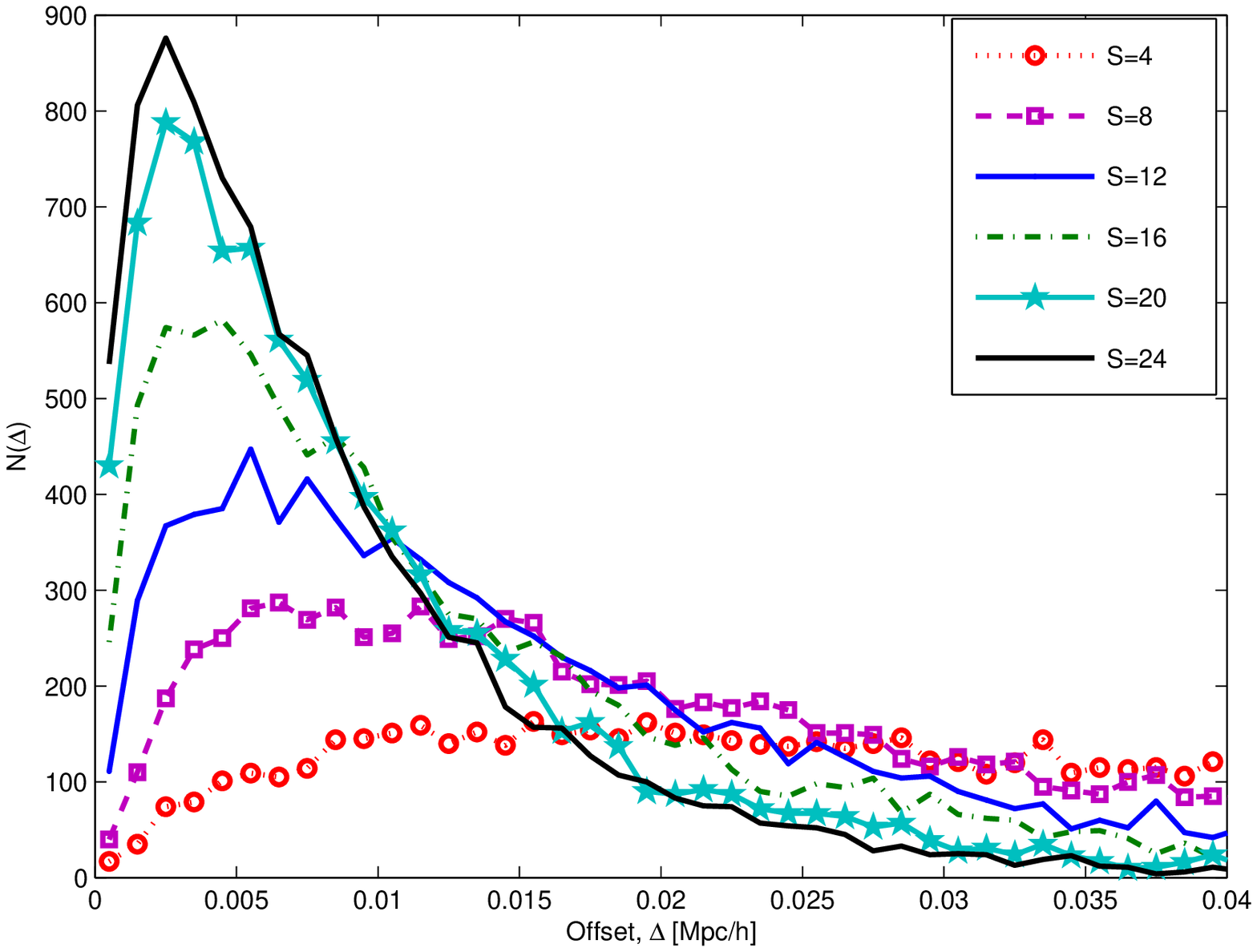}
\caption{Effect of the prior smoothing polynomial degree $S$ on the posterior distribution of offsets between the BCG assigned by the GMBCG catalogue and the DM centre (for fixed $q=1.2$). As higher $S$ values entail usually steeper mass profiles, and have more degrees of freedom to describe the substructure, higher $S$ values entail generally smaller BCG-DM offsets, converging towards the BCG near the boundary of our chosen range, $S=24$.}
\label{NvsS}
\end{figure}

\begin{figure}
\centering
 \includegraphics[width=90mm]{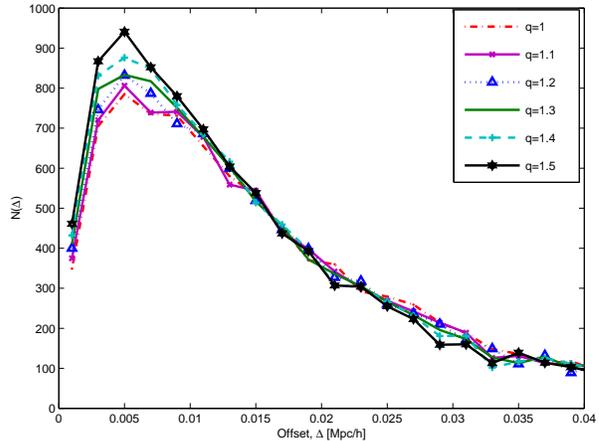}
\caption{Effect of the prior galaxy surface-density power-law $q$ on the posterior distribution of offsets (for fixed $S=12$). Compared to the effect of the polynomial degree seen in Figure \ref{NvsS}, the effect of the galaxy power law $q$ on the resulting distribution and the location of the peak, is negligible.}
\label{NvsQ}
\end{figure}

In a recent work, \citet{Einasto2012} studied substructure and multimodality in groups and clusters of galaxies in SDSS DR8, using several tests to characterise the member distribution. They found that the distribution of distances from the cluster (or component) centre for the corresponding brightest galaxies shows that most of these galaxies are located preferentially close to the (sub)cluster centre (typically of an order of $\sim0.1$ $h^{-1}$ Mpc scales, see Fig. 4 therein), although substantial numbers can show also higher separations (typically of a $\sim1$ $h^{-1}$ Mpc scale).

Here, we aim to characterise the typical offset between the DM peak and the BCG of a statistically-significant sample, with a successful mass-modelling tool for SL analyses \citep{Zitrin2009b} which does not require the DM centre to be predetermined, nor assumes a symmetric or any particular pre-known shape for the DM distribution. This method was recently adapted for automated use on 10,000 SDSS clusters, deducing the first observational, universal distribution of Einstein radii \citep{Zitrin2011d}. The advantage of this approach in evaluating the BCG offsets is that the resolution is very high (an order of 0.1\arcsec), and that the analysis can be performed blindly, on a very large sample. The results of this study will constitute the first statistically-significant, observational measure of this effect, and supply a complementary measure to compare to the previous studies mentioned above. In addition, the offset between the DM peak and the BCG may eventually help characterise the evolution of cluster galaxies and their host clusters, as more relaxed clusters can be anticipated to show decreasing or negligible offsets.

We analyse here the same 10,000 clusters from \citet{Zitrin2011d}, which were randomly drawn from the Gaussian Mixture Brightest Cluster Galaxy (GMBCG; \citealt{Hao2010GMBCG_cat}) SDSS cluster catalogue. In practice, these span the full redshift and richness ranges covered by the full catalog. The clusters had been found using the Error Corrected Gaussian Mixture Model algorithm \citep{Hao2009GMBCG_find} to identify the BCG plus red sequence feature, convolving the identified red sequence galaxies with a spatial smoothing kernel to measure the clustering strength of galaxies (within 0.5 Mpc) around BCGs. The technique was applied to the Data Release 7 of the Sloan Digital Sky Survey and produced a catalogue of over 55,000 rich galaxy clusters in the redshift range $0.1 < z < 0.55$. The catalogue is approximately volume limited up to redshift $z\sim0.4$ and shows high purity and completeness when tested against a mock catalogue, and when compared to other well-established SDSS cluster catalogues such as maxBCG (\citealt{Koester2007maxBCG_cat}; for more details see \citealt{Hao2010GMBCG_cat}).

\begin{figure}
\centering
 \includegraphics[width=80mm]{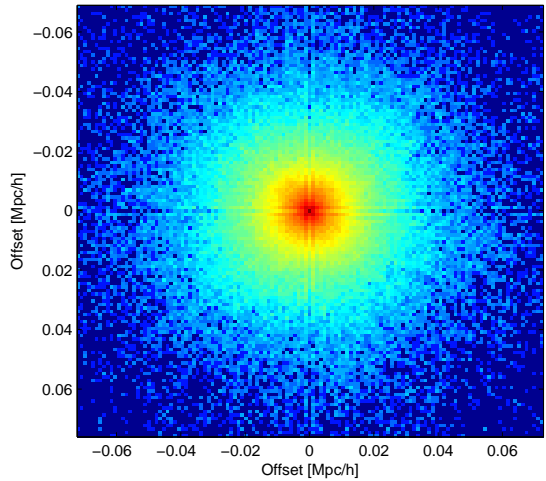}
\caption{2D (posterior) distribution of offsets between the BCG assigned by the GMBCG catalogue and the DM centre. As can be seen and expected, the offsets are distributed equally around zero with no preferred direction.}
\label{2dDist}
\end{figure}

The paper is arranged as follows: in \S 2 we detail the method incorporated in order to obtain the BCG-DM offsets. In \S 3 we report the results and factors of uncertainty, which are further discussed in \S 4 and concluded in \S 5. Throughout we use a standard $\Lambda$CDM cosmology with ($\Omega_{\rm m0}=0.3$, $\Omega_{\Lambda 0}=0.7$, $h=0.7$), and distances are usually given in $h^{-1}$ Mpc, to ease the comparison to other works. To avoid possible confusion, we also note that all logarithmic quantities and syntax in this work are in base 10, unless stated otherwise, and are denoted equivalently as either ``$\log_{10}$'' or ``\emph{Log}''.

\section{Method}\label{method}
Our analysis starts with the basic assumption that light traces mass. For each cluster, members were identified and listed in the GMBCG catalogue, following a photo-$z$, luminosity, and distance filters, and the red-sequence definition in the $g-r$ colour for $z_{l}<0.43$, or the $r-i$ colour for $z_{l}>0.43$ \citep[see][for more details]{Hao2010GMBCG_cat}. We assign listed members in the catalogue with a power-law surface density profile $q$, $\Sigma(r)\propto r^{-q}$ (where $r$ is the radius from the galaxy), scaled by their luminosity, so that the superposition of all galaxy contributions in the cluster field represents the galaxy ``lumpy'' component of the mass model. The lumpy mass distribution is then smoothed to obtain the DM distribution, where the smoothing is performed by fitting a (relatively) low-order polynomial using 2D spline interpolation. The polynomial degree, $S$, determines in practice the position of the DM centre, or peak, so that to cover all options, we must take into account all reasonable polynomial degrees (as explained shortly), and to sample a wide-enough range of power laws to verify this choice does not have an effect on the final result. Throughout, the DM centre is simply taken as the highest surface density peak of the DM map \citep[see also, e.g.][]{Navarro2010}.

\begin{figure}
\centering
 \includegraphics[width=95mm]{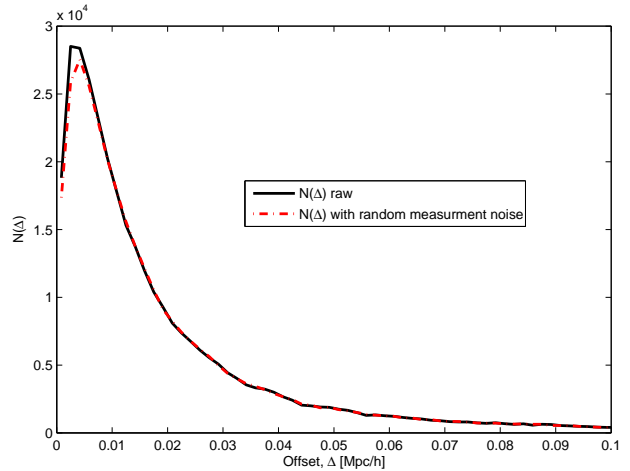}
\caption{Posterior distribution of offsets between the BCG assigned by the GMBCG catalogue and the DM centre obtained as described in \S \ref{method}. The \emph{black} curve shows the raw measurement, and the \emph{red} curve includes 1 SDSS pixel ($\simeq0.4\arcsec/pix$) random measurement noise, which we adopt hereafter for our analysis. The curve peaks at $\simeq0.0042$ $h^{-1}$ Mpc (or 1.3\arcsec if binned in arcseconds rather than physical distance), and with a median of 0.0136 $h^{-1}$ Mpc. Following previous work, we tried to fit a 2D Gaussian to the distribution, although this yields a poor fit of $R^{2}\sim0.7$ at best. The curve, however, is very-well described by a log-normal distribution, as we discover here and show in Figure \ref{1dDistLog}.}
\label{1dDist}
\end{figure}

This analysis is based on the lens-modelling method developed by \citet{Zitrin2009b}, adopting the approach used in \citet{Broadhurst2005a}, where for our purpose here we do not need to obtain the final (scaled) mass distribution, but only to use the first two steps: generating the lumpy and smooth (unscaled) components of the mass model. Clearly, the overall relative scaling of each component, obtained usually in a minimisation procedure using multiple-images, is irrelevant to the question of where the DM centre is, especially since the galaxy component is known to comprise anyhow only a small fraction of the total mass distribution \citep[see also][]{Zitrin2011d}.

Furthermore, we note that by performing the simple procedure described in \citet{Zitrin2009b}, namely, representing the red-sequence galaxies by a power-law surface density profile, smoothing it in order to obtain a representative DM component, and combining the two components with a reasonable and calibrated relative weight, we were able to immediately identify unprecedented numbers of multiple-images in many cluster fields. These multiple-images are physically found by this initial mass model, implying that this parametrisation is most successful in describing the underlying (projected) mass distribution. When performing a detailed mass modelling using multiple-images, one goes over a representative range of $q$ and $S$ values, and minimises according to the constraints comprised by the multiple-image locations and redshifts. In practice, these two parameters are controlling the overall mass profile (steeper power-laws and higher polynomial degrees usually entail steeper total mass profiles, see \citealt{Zitrin2009b}), which is constrained using the angular-diameter distance to the different multiple systems.

\begin{figure}
\centering
 \includegraphics[width=95mm]{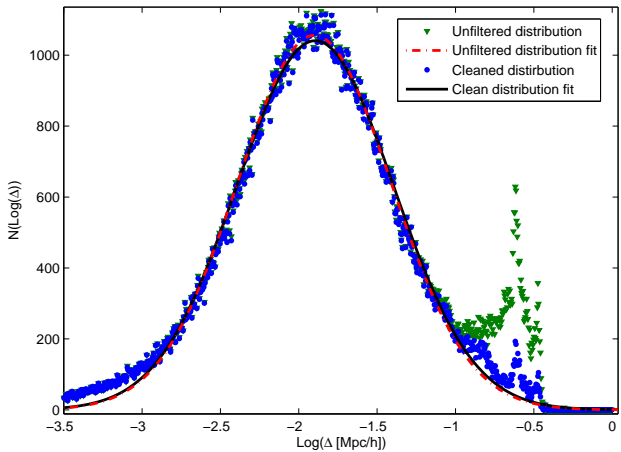}
\caption{Posterior logarithmic distribution of DM-to-BCG offsets. As is clearly seen, the offsets (\emph{green triangles}) are well described by a log-normal distribution, with $\langle \log_{10}(\Delta~[h^{-1}$Mpc$])\rangle=-1.898^{+0.002}_{-0.003}$, and $\sigma=0.489\pm0.004$ ($95\%$ confidence levels; $R^{2}=0.988$). The fit is shown in a \emph{dash-dotted red line}, done for all $\log_{10}(\Delta~[h^{-1}$Mpc$])<-1.11$ offset values thus excluding the artefact caused by our SL analysing frame size at higher offsets. This artefact is then filtered out by a simple noise cleaning procedure described in \S \ref{offsetNoiseS1}, and the result is overplotted in \emph{blue dots}. The resulting cleaned distribution maintains the same log-normal shape, width, and position, as can be seen by the \emph{solid black curve}, with $\langle \log_{10}(\Delta~[h^{-1}$Mpc$])\rangle=-1.895^{+0.003}_{-0.004}$, $\sigma=0.501\pm0.004$, and a slightly better $R^{2}$ of 0.992. This is fully consistent with the estimate by forcibly excluding outliers. Although similar, the latter fit, of the clean sample, is the final official result we adopt throughout.}
\label{1dDistLog}
\end{figure}

\begin{figure}
\centering
 \includegraphics[width=90mm]{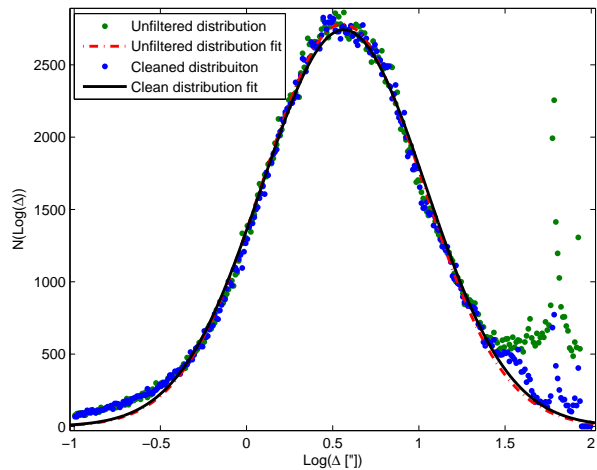}
\caption{Posterior logarithmic distribution of offsets, \emph{in arcseconds}, between the BCG and the DM centre. The offsets (\emph{green dots}) are well described by a log-normal distribution, with $\langle \log_{10}(\Delta~[\arcsec])\rangle=0.557\pm0.004$, and $\sigma=0.501\pm0.004$ ($95\%$ confidence levels; $R^{2}=0.995$). The fit, shown in a \emph{dash-dotted red line}, is performed while ignoring values below $\log_{10}(\Delta~[\arcsec])<1.3$, thus excluding the artefact caused by our SL analysing frame size at higher offsets. Similar to Figure \ref{1dDistLog}, this artefact is filtered out by a simple noise cleaning procedure described in \S \ref{offsetNoiseS1}, and the result is overplotted in \emph{blue dots}. The resulting cleaned distribution maintains the same log-normal shape, width, and position: $\langle \log_{10}(\Delta~[\arcsec])\rangle=0.564\pm0.005$, $\sigma=0.475\pm0.007$, and $R^{2}=0.99$, clearly consistent with the result obtained prior to the cleaning procedure by excluding outliers.}
\label{1dDistLogarcsec}
\end{figure}

Here, we do not use multiple images as constraints. The large numbers of multiple-images found by the mass models constructed following the simple procedure described here, in previous clusters we analysed, use as an evidence and show that this method describes with unprecedented reliability the underlying mass distribution. In other words, the $\sim30$ clusters we accurately analysed with this method to date \citep[e.g.][]{Broadhurst2005a,Zitrin2011a,Zitrin2011b,Zitrin2011c}, show unambiguously that the $q$ and $S$ parameter space in the relevant range contains a representative solution for the cluster mass distribution, and hence, the DM centre. The polynomial degree, $S$, controls also the effective centre of the DM component, for a given power-law, $q$. Here, as mentioned, since we do not use multiple-images as constraints, we go over a full range of polynomial degrees to account for all possible positions of the DM centre, probing also different power-law values to test their influence on the resulting DM peak position. The realistic ranges and values we adopt are [$1\leq q \leq 1.5$] with $\Delta q=0.1$, and [$4\leq s \leq 24$] with $\Delta S=4$, so that the $q-S$ grid comprises 36 measurements per cluster. This prior distribution we adopt is flat, so that each point on the $q-S$ grid is equally weighted. This range also contains the parameter values obtained in the minimisation procedure when performing detailed SL analyses of many cluster lenses mentioned above. Moreover, a higher polynomial degree than the range we adopt here does not affect the DM centre significantly, i.e., the DM peak converges towards the BCG centre with higher polynomial degrees, so that above $s\sim20$ the DM peak will usually approach the BCG, as there are enough degrees of freedom to describe smaller substructures (less smooth a fit). Not to create a bias towards smaller offsets, we therefore do not use smoothing degrees above $s=24$. See Figures \ref{NvsS} and \ref{NvsQ} for the effect of the $q$ and $S$ parameters on the posterior distribution of BCG-DM offsets (this is discussed further in \S \ref{systematic}). This in fact leads to the key point of our analysis: we sample the full range of possible DM centre positions per cluster, which are determined by the parametrisation we described, and entailed from the original galaxy distribution. In that sense, the DM centre is approximated by the weighted centre of luminosity, which is also affected therefore by the luminosity density and not only the brightest single member: clusters with evenly distributed galaxy luminosity around the BCG are more likely to show very small range of offsets regardless of the polynomial degree, while clusters with concentrated luminous and massive group(s) close to the BCG may show a wider range of possible DM-BCG offsets, strongly depending on the polynomial degree.

Throughout we work in a field-of-view of $120\arcsec\times120\arcsec$, since this is typically a large-enough size to describe the SL field. For example the largest Einstein radius known to date, see \citet{Zitrin2009a}, is $\sim55\arcsec$ for redshift of $z\simeq3$ \citep{Limousin2011_0717}, and since the critical curve is approximately centred on the centre-of-mass, it is clearly not physically likely that the true offset would be significantly larger than this (except for, say, clusters with other galaxies which are indeed brighter than the massive galaxy sitting at, or close to, the potential minimum, see \S \ref{offsetNoiseS1}). Also, from previous work we know that usually lensing clusters show up to a $\sim5\arcsec$ offset or less (although some may show a significantly higher offset, see below), as revealed in detailed HST-based SL analyses (e.g., \citealt{Zitrin2009a,Zitrin2009b,Umetsu2011}, see also \citealt{Smith2005,Richard2010locuss20}). This is indirectly also supported by the result of \citet{Johnston2007offset} who found that misidentified clusters in a similar SDSS catalogue, show a 2D Gaussian distribution with $\sigma=0.42 h^{-1}$ Mpc, and that correctly identified clusters should comprise offsets much closer to the BCG (as seen also in the other studies mentioned in \S \ref{intro}; e.g. \citealt{Oguri2010_25clusters,Shan2010_38offsets,George2012offsets}), which we have taken upon to characterise here. In that context, it is worth mentioning that the method we apply here is capable of tracing also larger offsets than few arcseconds (if entailed by, e.g., the location of multiple images), and thus can probe a large enough offset range. One such example is the largest Einstein radius cluster mentioned above, Macs J0717.5+3745, with an offset of $\sim30\arcsec$ measured from our model presented in \citet{Zitrin2009a}. However we also note, that were there correctly identified clusters with significantly larger offsets than we probe in our work (e.g., much further away from of the SL region or the FOV we work in), these might not be uncovered by our method, although we conclude from the above considerations, that these cannot be expected to be significant.

The procedure implemented here can be thought of as simply smoothing the red sequence member light distribution, in order to obtain the range of possible DM centres. As simple as it may sound, the success of the light-traces-mass assumption in constantly describing the mass distribution in many lensing clusters, and the large statistical sample which helps to reduce the statistical uncertainties, aid us to obtain a well-constrained solution for the distribution of BCG-DM offsets, as seen in the following section (\S \ref{results}).

\section{Results and Uncertainties}\label{results}

In this section we present the results obtained by performing the analysis described in \S \ref{method}, assess the level of noise and uncertainty in them, and describe the noise-cleaning procedures taken.

\subsection{Offset Distribution}\label{offsetdistS1}
We examine the 2D distribution of BCG-DM offsets obtained in our analysis. In Fig. \ref{2dDist}, we show the 2D distribution of offsets for the 10,000 clusters analysed in this work. As can be seen and expected, the offsets are equally distributed around the centre in all directions, with no particular direction preference. Our first step, is to try and fit a 2D Gaussian to these data, $P(R_{s})=\frac{R_{s}}{\sigma^{2}} \exp(-\frac{1}{2} (\frac{R_{s}}{\sigma})^{2})$, motivated by the works of \citet{Johnston2007offset} and \citet{Oguri2010_25clusters}, where $R_{s}$ is the offset size (which we denote here throughout equivalently as $\Delta$). Figure \ref{1dDist} shows the probability density, or the number of clusters per (radial) offset bin, where we add therein the same distribution obtained by taking into account random noise of blending due to the SDSS relatively large pixel scale, $\simeq0.4\arcsec/pix$, although its effect on the location of the peak is negligible as seen. The peak of the distribution is very small, 0.0042 $h^{-1}$ Mpc, but significantly larger than zero, which is one of the important results of this work as we discuss below (\S \ref{discussion}).

No 2D Gaussian is capable of describing well the observed shape of the distribution. The best obtained 2D Gaussian fit has a poor goodness-of-fit estimator of $R^{2}\sim0.7$ at best. We also note, that we repeated the procedure in \emph{arcsecond} bins rather than $h^{-1}$~Mpc bins. The resulting histogram peaks at 1.3\arcsec, has the exact shape as the histogram seen in Figure \ref{1dDist}, and the same inconsistency with a 2D Gaussian fit.

We now remake the histogram, or probability distribution of offsets, in logarithmic bins. The logarithmic distributions are plotted in Figures \ref{1dDistLog} and \ref{1dDistLogarcsec}, and are clearly well described by log-normal distributions (of the form $f(\log_{10}(x))\propto\exp(\frac{-(\log_{10}(x)-\langle \log_{10}(x)\rangle)^2}{2\sigma^2})$), with $\langle \log_{10}(\Delta~[h^{-1} $Mpc$])\rangle=-1.898^{+0.002}_{-0.003}$, and $\sigma=0.489\pm0.004$ ($95\%$ confidence levels; $R^{2}=0.988$), and $\langle \log_{10}(\Delta~[\arcsec])\rangle=0.557\pm0.004$, $\sigma=0.501\pm0.004$, respectively. Outliers, defined as $\log_{10}(\Delta~[h^{-1} $Mpc$])>-1.11$, or $\log_{10}(\Delta~[\arcsec])>1.3$, are excluded from the fit, as we now describe (\S \ref{offsetNoiseS1}).

\subsection{Noise and Uncertainty}\label{offsetNoiseS1}
\begin{figure*}
\centering
 \includegraphics[width=140mm,height=100mm]{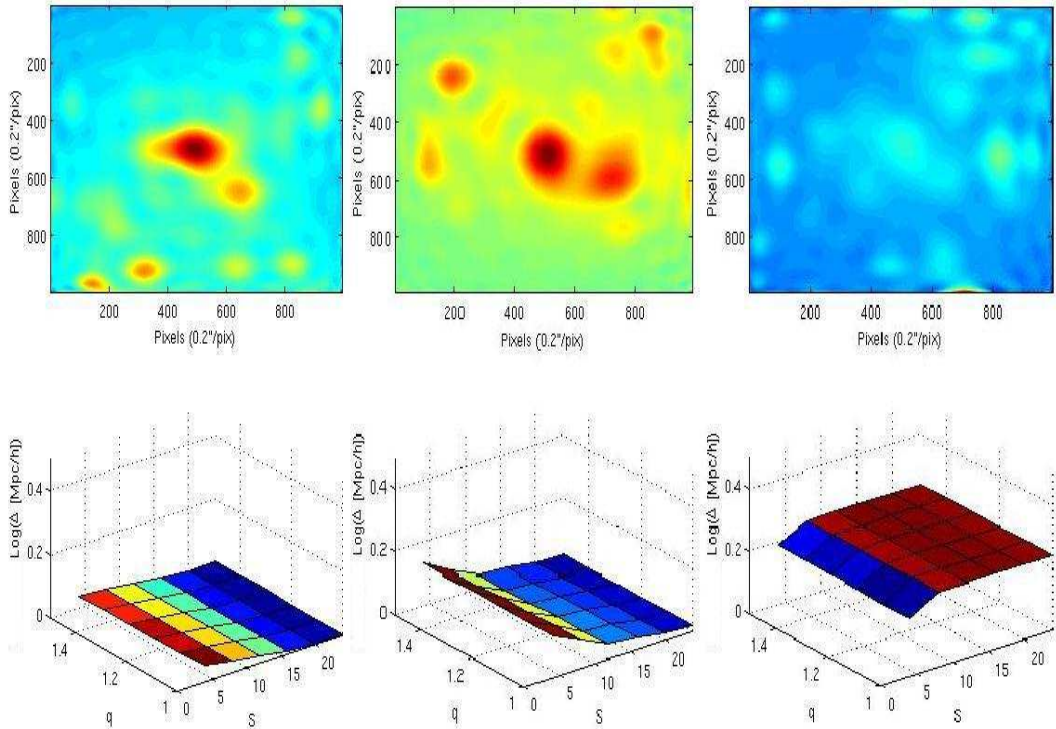}
\caption{Example of three different scenarios seen in a check by eye of 300 randomly selected clusters, that helped to filter out most of the noise of spurious BCG or cluster detections. For each scenario the \emph{upper panel} shows the resulting DM distribution, and the \emph{lower panel} shows the offset deduced for each cluster in the $q$ and $S$ space. \emph{Left:} A typically relaxed cluster, with one main central halo and other smaller and less massive mini-halos. This cluster is correctly identified and the offset values are in the range physically expected for relaxed clusters. \emph{Centre:} A less relaxed, probably merging cluster. This cluster shows more prominent subhalos close to the central BCG, and the values for the offsets in the $q$ and $S$ grid follow the expected trend, and although partially large - we consider such clusters, or offsets, possibly real and include them in our analysis. \emph{Right:} A misidentified cluster. No prominent BCG halo is seen in the DM map, and, the offset values in the $q$-$S$ plane do not follow the expected relation and lie all above a higher threshold. We therefore make use of this quality to filter out such clusters from our further analysis. For more details see \S \ref{offsetNoiseS1}.}
\label{examples3}
\end{figure*}

In Figures \ref{1dDistLog} and \ref{1dDistLogarcsec}, a significant secondary peak is seen at $\log_{10}(\Delta~[h^{-1}$Mpc$])\sim-0.6$, or $\log_{10}(\Delta~[\arcsec])\sim1.8$, which deviates from the log-normal shape that characterises the distribution at lower offsets. The origin of this peak is quite clear, as it forms around the edge of the field-of-view (FOV) size we use for our lens modelling, of $120\arcsec\times120\arcsec$, or an effective radius of $\sim60-85\arcsec$ (\S \ref{method}), so that clusters exhibiting (unrealistic) offsets larger than this, are counted at that maximal offset range. This range translates into $\sim0.078-0.11~h^{-1}$ Mpc for the redshift range of the sample, $0.1<z<0.55$, or covering the logarithmic offset range of $-1.11<\log_{10}(\Delta~[h^{-1}$Mpc$])<-0.42$, exactly where the peak is seen. The first step we therefore perform, is compare the integral of the log-normal distribution fit to the main peak in the range of the secondary peak, to the area enclosed below the data themselves. We obtain that the area enclosed by the data in that range, is $\sim2.5$ times the area expected by the fit, implying a noise level of $\sim150\%$ in that particular ``problematic'' range. We find that about 1250 clusters out of the 10,000 analysed contributed to this secondary peak (or close to 50,000 out of 360,000 individual measurements), where only $\sim500$ are expected by the fit. Since true (rather than misidentified) clusters are not expected to have such large offsets, this fraction we adopt as the level of misidentified clusters in the catalogue, roughly $(1250-500)/10000\sim8\%$. We also address the reader to Figure \ref{examples3}, where we show different examples of clusters probed by our method, with a clear deviation for misidentified clusters. Clearly, there could be other misidentified clusters or BCGs overlooked by this procedure, that contribute to the smaller offset range inputting some additional noise to the offset distribution. Previous studies \citep{LinMohr2004,Johnston2007offset,Koester2007maxBCG_cat,Oguri2010_25clusters,Hao2010GMBCG_cat,Andreon2011darkclusters,Skibba2011centralBCG,George2012offsets} showed that usually an order of $\sim10-30\%$ of the clusters in such catalogues are misidentified, so that after accounting for the misidentified clusters which affect the high-end offsets, the remaining number of misidentified clusters in the sample can be expected to be reasonable (note also that some of the contribution of the same misidentified clusters to the small offset range was automatically excluded when filtering out problematic clusters as above). The overall level of misidentified clusters in the sample can be therefore expected to be around $\sim10-20\%$, or even up to a higher level of around $\sim30\%$ following previous works \citep[e.g.][]{Johnston2007offset}. Remaining discrepancy from the $\sim10\%$ level we find, may be attributed to the amount of misidentified clusters that are \emph{not} traced as such by our method, thus implying a possible $\sim10-20\%$ (additional) noise level on our results from such clusters.

The second step, is to search whether there exists a certain population of clusters which contributes specifically to this artefact, and later, verify that by excluding these clusters, the log-normal fit to the true lower offsets, obtained in \S \ref{offsetdistS1}, remains consistent. We therefore examine various properties as a function of $\log_{10}(\Delta)$, looking for distinguishable populations. For example, we probed if the $\log_{10}(\Delta~[h^{-1} $Mpc$])>-1.11$ points originate from a certain redshift, richness, or relaxation-degree populations. No specific distinguishable population (in terms of relative number of misidentified clusters per criterion bin) is uncovered as a function of richness or redshift, and clusters from all redshift or richness bins contribute to the artefact (these are therefore not shown here). In order to examine the relation with relaxation degree, we examined by eye 300 randomly selected clusters, 30 clusters in each of the ten $\log_{10}(\Delta~[h^{-1} $Mpc$])=0.5$ wide bins (in the range [$-5\leq\log_{10}(\Delta~[h^{-1} $Mpc$])\leq0$]), where for each cluster we marked down whether it seems relaxed (e.g., if the mass/light is clearly concentrated around the BCG identified by the catalogue), or not necessarily. Figure \ref{examples3} (\emph{upper panel}) shows an example of three such distinctive cases. In addition, for each cluster probed by eye, we examine also the distribution of offsets in the $q$ and $S$ grid (Figure \ref{examples3}, \emph{lower panel}). This, since the check by eye is highly subjective, and the shape of offsets on the $q$ and $S$ grid may withhold the information needed to filter this artefact more quantitatively.

We noticed, as can be expected, that unrelaxed clusters tend to contribute more to the artefact than relaxed clusters, but that distinction is weak and subjective, and we do not make further use of it in our work. However, as suggested above, by performing this procedure, we found that indeed filtering based on the offset distribution per cluster, in the $q$ and $S$ grid, helps to remove most of the artefact noise. As can be seen in Figure \ref{examples3}, clusters of the third kind therein (\emph{right column}) exhibit offsets which are significantly larger than well-identified clusters, all throughout the $q$ and $S$ plane. Since one of our assumptions is that higher $q$, but especially $S$ values, should comprise, in general, smaller offsets converging towards the BCG, the fact that in these clusters there is no such trend and the whole plane suggests high separation, seems to represent well the misidentified portion of clusters (or BCGs). In practice, by several iterations, we found that the \emph{median} offset per cluster in the 36-point $q$ and $S$ plane is a good measure for this deviation, which should help in the noise/artefact cleaning procedure. Excluding clusters with median($\Delta$)$>40\arcsec$ (or any other threshold value somewhat lower than this), preserves the shape and location of the original main peak, and gets rid of most of the noise: about 600 clusters out of the $\sim750$ noise clusters were filtered out. This means that the overall noise level is now down to $150/500\sim30\%$ in that region. This noise level is low enough to now reinclude that noisy region in the fit, so the fit will now be obtained by the entire range, and the (remaining) noise level in the high-end region will play only an insignificant role.

As a last verifying step, we analysed a sub-sample of a few dozen clusters chosen randomly from the offsetted, misidentified candidate clusters, with a larger FOV, to make sure the noise was indeed an artefact from the frame size. Most of these clusters, as expected, created now a peak around the new FOV size. We note also, that analysing the full sample with a larger FOV is too expensive computationally, and in fact due to the reasons we elaborated here, there is actually no need to do so (eventually that would just shift the artefact peak further away).

In Figures \ref{1dDistLog} and \ref{1dDistLogarcsec} we replot the probability density, or logarithmic histogram of offsets, after excluding the clusters designated as misidentified using the above criterion. The secondary peak, as seen, almost vanishes, and the overall fit to the entire range, is $\langle \log_{10}(\Delta~[h^{-1}$Mpc$])\rangle=-1.895^{+0.003}_{-0.004}$, $\sigma=0.501\pm0.004$ and $R^{2}$ of 0.992 (or $\langle \log_{10}(\Delta~[\arcsec])\rangle=0.564\pm0.005$, $\sigma=0.475\pm0.007$ and $R^{2}=0.99$ for the arcsecond logarithmic distribution), fully consistent with the result obtained above excluding high-offsets form the fit. This verifies our noise cleaning procedure efficient and secure, and the result of this fit is the one adopted as the final shape of BCG-to-DM offsets found in this work.

Due to the very large number of clusters analysed, statistical uncertainties could in fact be expected to be very small. One of the possible factors of uncertainty, is the effect of the photo-$z$ uncertainty on the distribution in physical scales ($h^{-1}$ Mpc, Figure \ref{1dDistLog}), and should not affect the arcsecond distribution (Figure \ref{1dDistLogarcsec}), where there is no conversion to distance. The typical photo-$z$ uncertainty for the sample BCGs is 0.015. Assuming this error is relatively symmetric around the real redshift, the overall statistics of offsets cannot be expected to change significantly. To assess the level of uncertainty the photo-$z$ error may contribute, we repeated the analysis over the full 10,000 cluster sample by photometric redshifts drawn randomly from a normal distribution centred on the catalogue photometric redshift for each cluster, with a width of $\sigma=0.015$ (which is the photo-$z$ error quoted in Hao et al. 2010). We obtain that the distribution remains nearly unaffected: $\langle \log_{10}(\Delta~[h^{-1}$Mpc$])\rangle$ and $\sigma$ change by less than $0.1\%$.

Another factor of possible uncertainty is the portion of cluster galaxies correctly assigned by the cluster finding algorithm and used to construct the mass model. Since we work with the assumption that light traces mass, our method is strongly coupled to the input cluster members and their photometry. For example, some clusters may exhibit galaxies which are brighter in practice than the massive galaxy sitting at, or close to, the cluster potential well. In such cases, the catalogue is likely to assign the brighter galaxy as the ``central'' one, and a large offset will be measured to the true, physical centre. In addition, were (other) members erroneously assigned to a cluster (including different structures along the line-of-sight), or true members went uncovered, this should have an effect on the resulting DM centre. Luckily, brighter members are more probable to be correctly assigned for a given cluster, which are those governing the fit. In addition, the effect of the DM polynomial fit degree is the most affecting parameter (Figure \ref{NvsS}), and since for each cluster we take into account a wide range of polynomial degrees resulting in a wide range of possible DM centres, the relative effect of misidentified cluster members should be comparably minor. To test this quantitatively we resampled 100 randomly chosen clusters from the catalogue, where for each cluster we input additional (artificial) cluster members in random locations in the frame, and compare the result to the analysis before including these. To construct a strong upper limit, for each cluster we input $50\%$ additional members than the current number of members listed in the catalogue, with brightnesses (and masses) random up to that of the BCG. By doing so we deduce that this inclusion of up to $50\%$ more members changes the offsets by typically, an order of $0.1\arcsec$ (and in any case less than $2.2\arcsec$), per cluster, resulting in less than a $0.1\%$ change on the mean and width of the distribution for these clusters. Overall 10,000 clusters, this effect is clearly negligible.

In addition, we note that we examined the effect of the richness of the clusters, specifically the number of members in each cluster, on the distribution. Such a trend was found, for example, by \citet{Johnston2007offset} for the significantly offsetted or misidentified clusters. We find no such noticeable trend, so clusters from all richness ranges contribute equally to the true offsets.

Although these may be hard to assess quantitatively, some additional contaminating factors should be at least acknowledged. For example, we note that per cluster, both the BCG and correspondingly, the DM centre deduced, originate from the same cluster catalogue. Therefore, were the true BCG not assigned to the cluster (often because its colours are bluer that other red-sequence galaxies; e.g., \citealt{Bildfell2008misBCG,Pipino2011misBCG}), clearly it would be centred on the wrong bright galaxy which we consider accordingly to be the BCG. This would render the offsets for this cluster non-credible. The level of such ``bluer'' BCGs which may be uncovered was previously assessed by \citet{Pipino2011misBCG}, for example, to be only $\sim8$ per cent in a 69,000 SDSS cluster catalogue, and so its effect on the whole sample should be of the same order. Interestingly, in a smaller sample of 48 X-ray luminous clusters, \citet{Bildfell2008misBCG} found a higher percentage of bluer-core BCGs ($\sim25\%$), and noted that in these clusters, the BCG lies within $\sim10$ kpc of the X-ray peak.

It should be stressed, in addition, that we do not attempt to deduce one \emph{single} offset measurement per cluster, but for each cluster expand our full parameters space to include a full \emph{range} of possible offsets (coupled to our prior distribution based on the success of this modelling method). The high number of clusters analysed in this manner allow us to deduce the credible posterior distribution of offsets presented here, given the prior choice of parameters range, as we now discuss.

\begin{figure*}
\centering
 \includegraphics[width=180mm]{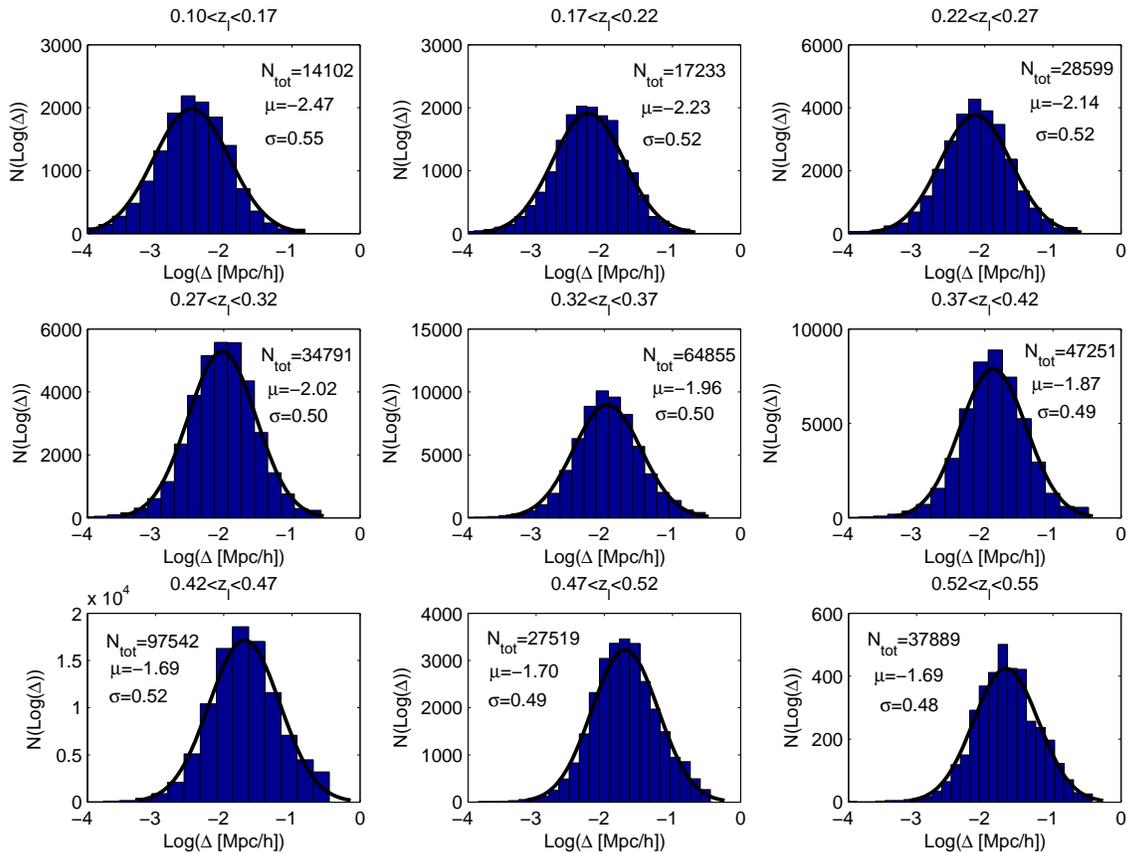}
\caption{Offset histograms as a function of redshift. For each bin we denote the number of data points, $N_{tot}$, and the mean and width of the distribution, $\mu$ and $\sigma$, respectively. Although insignificant given the distribution width in each bin, a steady trend is seen in the mean of the sample, where higher redshift clusters show higher BCG-DM separations, and generally, lower widths of the distribution (although the high-end of the last three bins may be affect by the size limit of our analysing frame). See also Figure \ref{growthMPC}.}
\label{histvsz}
\end{figure*}

\begin{figure*}
\centering
 \includegraphics[width=180mm]{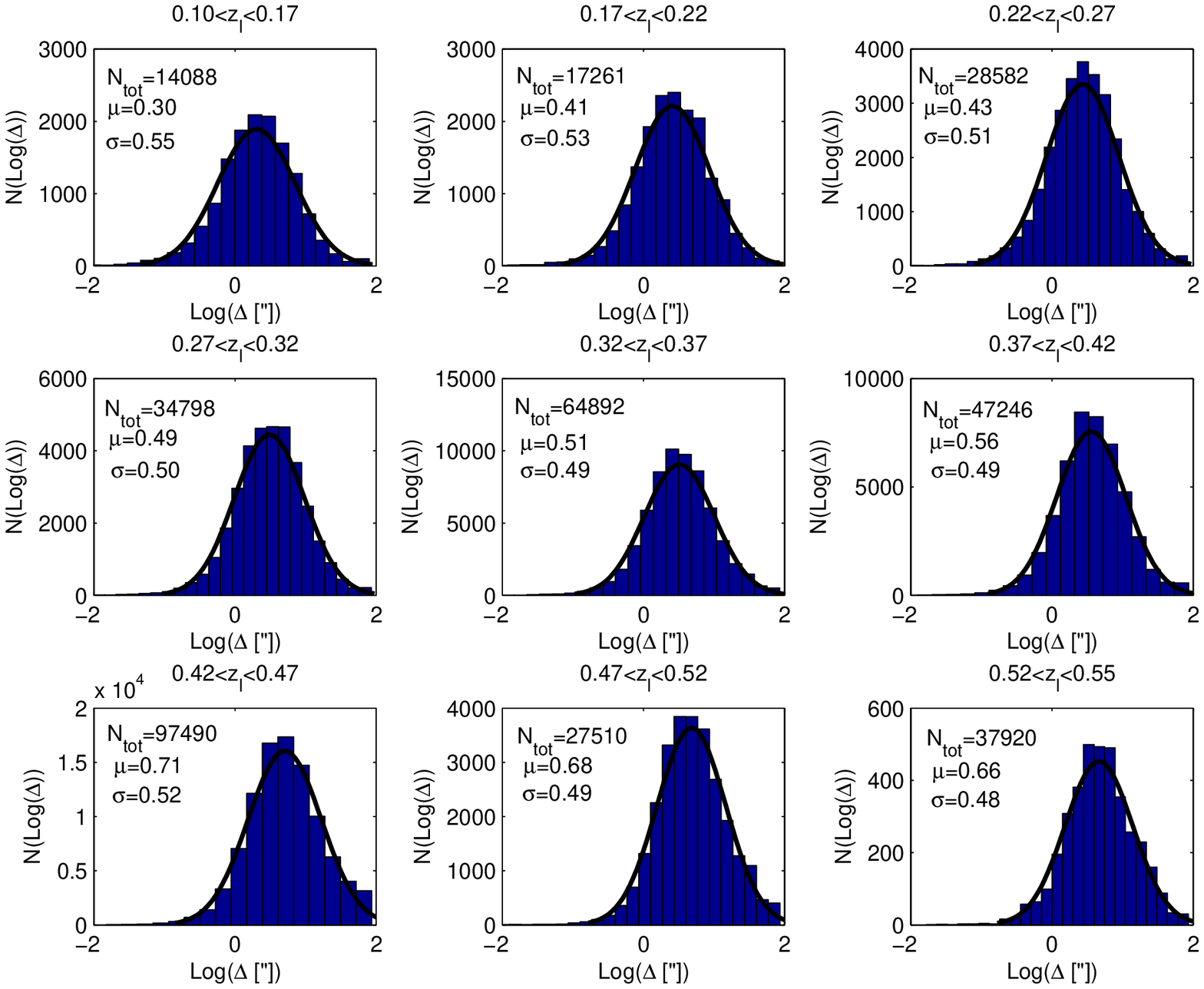}
\caption{Same as Figure \ref{histvsz}, but for distributions in angular logarithmic scales. See also Figure \ref{growthArcsec}.}
\label{histvszarcsec}
\end{figure*}

\subsection{Systematic Uncertainty from Priors}\label{systematic}

In the previous subsection, we discussed some (statistical and systematic) uncertainties relevant for our result. Here, we now aim to discuss the effect of prior choice of parameters on the resulting distributions. As seen in Figures \ref{NvsS} and \ref{NvsQ}, the resulting posterior distribution depends insignificantly on the prior distributions adopted for the $q$ parameter, but strongly depends on the $S$ parameter. Since fits are not constrained by multiple images, for each cluster one does not know what the best-fitting $q$ and $S$ values would be (which would result in a single offset value per cluster). The only way to quantify related uncertainties is to probe a broad range of $q$ and $S$ parameter values per cluster. We choose $q\in[1,1.5]$ and $S\in[4,24]$ (see \S \ref{method}) and sample this parameter range at 36 equidistant points per cluster. These samples are used to construct the 360,000 point histograms seen throughout. This amounts to Monte-Carlo sampling with wide, flat priors for $q$ and $S$ and takes into account the large uncertainties arising from the possible range of parameters. The resulting offset distribution is convolved with the (inferred) intrinsic distribution with the posterior distribution corresponding to Figures \ref{NvsS} and \ref{NvsQ}. Flat priors are chosen primarily in order not to bias the results by the lack of knowledge on how the $q$ and $S$ parameters should be distributed in practice.

The first step in assessing the resulting level of systematic uncertainty is therefore simply to examine Figures \ref{NvsS} and \ref{NvsQ}. The boundary cases (e.g., between $S=4$ and $S=24$) comprise a maximal systematic uncertainty of (the order of) $\log_{10}(\Delta~[h^{-1} $Mpc$])=-0.45$, for example, about $\simeq24\%$ of the obtained mean offset ($\langle \log_{10}(\Delta~[h^{-1} $Mpc$])\rangle=-1.895$). However, a more realistic estimation of the level of systematic uncertainty may come from incorporating more specific prior distributions for the $q$ and $S$ parameters. Since we have analysed about 30 clusters with our method previously, we can make use of the $q$ and $S$ distribution among these clusters to construct more sophistically guessed priors. Although these are not shown here, they entail a distribution of, \emph{crudely}, mean of $q\simeq1.25$ and $S\simeq10$, with $\sigma_{q}\simeq0.15$ and $\sigma_{S}\simeq4$. We use now these priors to generate a posterior distribution to compare with the posterior distribution obtained by initially incorporating flat priors. By doing so we obtain a systematic uncertainty of $\sim15\%$ on the log-normal mean, and $\sim10\%$ on the width of the (log-normal) distribution.

Throughout, however, we adopt and work with the result of the flat priors, since the allegedly more specific priors are based on previous analyses of HST images. One has to construct a (larger) sample of the SDSS clusters well-constrained with multiple images \citep[e.g.][]{Hennawi2008arcs,Bayliss2011,Oguri201238clusters}, in order to securely adopt an interior, prior distribution for $q$ and $S$. Until an explicit analysis of such a sample with our method is made available, we therefore adopt the more conservative, flat prior distribution results.

\section{Discussion}\label{discussion}

Although only relatively little work was conducted on the subject, the DM to BCG offsets have been increasingly examined in recent years, due to growing technical and observational capabilities enabling statistical studies of large samples, such as stacked WL analyses, and since a knowledge of the DM-BCG typical offset plays an important role in them. Here we further discuss the results of our work and other possible effects of uncertainty.

In \S \ref{results}, and Figures \ref{1dDistLog} and \ref{1dDistLogarcsec}, we presented the measured distribution of BCG-DM offsets. An important question that arises given the measured distribution, is whether the BCGs are on average centred in their cluster
potentials (described by the smoothed light distribution) with some finite deviation, or whether there is any
evidence for a significant offset. To address this question, we
perform a student-t test, which tests the hypothesis that a set
of random variables is drawn from a Gaussian with either known
or unknown mean and variance. The specific hypothesis tested in
our case is whether the measured offsets can be considered as drawn
from a Gaussian with zero mean and arbitrary variance. Suppose
the projected BCG coordinates relative to their cluster centres
are indeed random variates drawn from a two-dimensional
Gaussian distribution, then their radial distances to the
cluster centres, which we have measured here, are clearly not
distributed in a Gaussian way. However, we can then return the
radial distances into a set of Gaussian random variates by
multiplying with the sine or cosine of a random phase with a
flat distribution between $[0, 2\pi]$. This multiplication with a random phase removes the geometrical effect that the differential
area shrinks proportional to the distance from the centre, which makes radial distributions
peak at finite radii. This corresponds to
projecting the radial distance on an arbitrarily oriented
coordinate axis. Thus, if our hypothesis is true, the radial
distances multiplied with the sine or cosine of a random phase
angle are Gaussian variates $\{x_i\}$ with zero mean. To this
set of numbers, the student-t test can now be directly applied.

Its normalised test statistic,
\begin{equation}
T(x) = \frac{\sqrt{N}\bar x}{s(x)}\;,
\end{equation}
where $\bar x$ and $s(x)$ are the mean and the standard deviation of the set
$\{x_i\}$, follows a student-t distribution with $N-1$ degrees
of freedom if the hypothesis is true. For our sample, we
clearly have to reject this hypothesis. While the set $\{x_i\}$
can be well described by a Gaussian distribution, this
distribution does not have zero mean. We obtain that the absolute t-statistics is 0.1(0.19), with an error probability of 0.42(0.46), that the linear(logarithmic) distribution is drawn from a Gaussian with a zero mean. We therefore conclude that although relatively small, the typical offset is significantly, non-zero.

\begin{figure}
\centering
 \includegraphics[width=90mm]{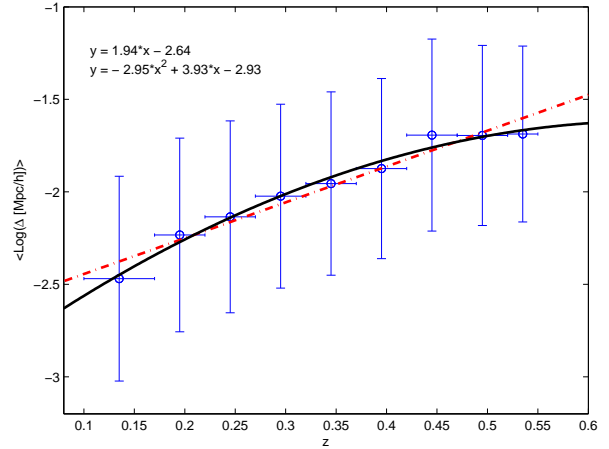}
\caption{Growth of the log-normal distribution mean, $\langle \log_{10}(\Delta [h^{-1} $Mpc$])\rangle$ (\emph{open circles}), and width, $\sigma$ (\emph{vertical error bars}), as a function of redshift. The \emph{horizontal error bars} represent the redshift bin width. Although it is insignificant, i.e. of the order of $1\sigma$, a tentative trend is seen as a function of redshift, so that higher separations and lower widths are seen for higher-$z$ clusters. The linear and quadratic least-squares fit to the data are given in the upper left corner, and overplotted as dash-dotted red and black lines, respectively.}
\label{growthMPC}
\end{figure}

\begin{figure}
\centering
 \includegraphics[width=90mm]{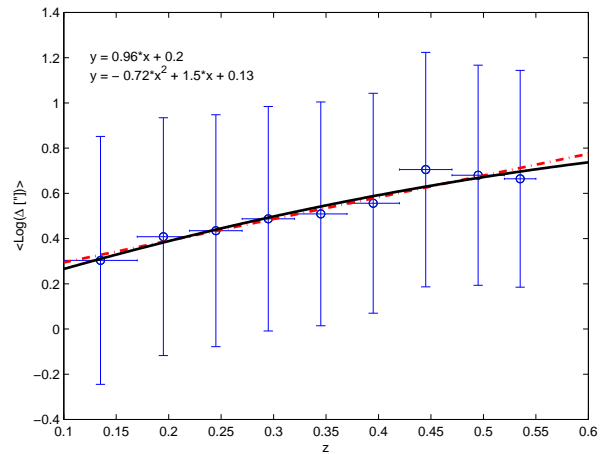}
\caption{Same as Figure \ref{growthMPC}, but for the log-normal distribution in arcsecond binning. This Figure, along with Figure \ref{growthMPC}, shows that there exists a tentative (of the order of only half a $\sigma$ here) but steady trend possibly originating from the evolution and merger history of clusters, so that higher redshift and therefore less relaxed clusters tend to show (both physical-distance and angular) higher separations, and lower (log-normal) distribution widths. One way to interpret the minor distribution-width trend, is that higher redshift clusters exhibit a somewhat more coherent population, say, in terms of relaxation, than lower redshift clusters
that show a wider variety of relaxed and unrelaxed clusters.}
\label{growthArcsec}
\end{figure}

In Figures \ref{histvsz} and \ref{histvszarcsec} we plotted the offset distribution in different redshift bins. The evolution of the mean and width of the (log-normal) distributions as a function of bin redshift is summarised in Figures \ref{growthMPC} and \ref{growthArcsec}. As can be seen therein, the mean of both the (log-normal) distribution in physical scales and in angular scales, increases steadily with redshift. The observed evolution, is however not significant: it is of the order of $1\sigma$ for physical scale distribution, and only half a $\sigma$ for the angular scale offset distribution. This low significance renders these trends at best, tentative. We note, however, that in a recent work \citep{Zitrin2011d}, in which we analysed the Einstein radius distribution of the same GMBCG catalogue, we also uncovered similar (insignificant but monotonic) evolvement in redshift: throughout the same (volume-limited) redshift range, the mean Einstein radius decreases continuously with redshift. If real, these evolvements in redshift, in both the Einstein radius and the DM-BCG offset distributions, could together help characterise the evolution, relaxation, and merger history of galaxy clusters more generally, in addition to other complementary studies.

As mentioned (\S \ref{intro}), \citet{Oguri2010_25clusters} observationally examined the offset of the BCG from the centre of mass obtained in weak lensing (WL) analyses of a sample of 25 clusters. They found that the DM centre is overall consistent with that of the BCG (within $2\sigma$ level), and that the observed distribution can be described by two components. The first, significant component describing the small offsets, is a 2D Gaussian with $\sigma=0.09~ h^{-1}$ Mpc, and the second less-significant component describing the tail of larger separations, is fitted by the \citet{Johnston2007offset} finding \citep[see also][]{HilbertWhite2010}: a 2D Gaussian with $\sigma=0.42~ h^{-1}$ Mpc. Note that in our work we do not characterise the offset distribution of the misidentified clusters, but only the correctly identified ones. In addition, we cannot explicitly compare to the results of \citet{Oguri2010_25clusters}, since in their work the centres are those determined by fitting a symmetric DM distribution (e.g., NFW) to the overall, larger-scale WL data (this in fact was recently found to be problematic especially in merging clusters, see \citealt{George2012offsets}), while in our work we simply measure the range of possible locations of the central DM peak.

\citet{Shan2010_38offsets} characterised the offsets between the X-ray peaks and lensing centres in 38 clusters (see also \citealt{Allen1998L-XrayDisc}). Although most clusters show small offsets, as is also usually seen in such lensing-centre to BCG comparison \citep[e.g.][]{Smith2005,Richard2010locuss20} showing typically less than $5\arcsec$ offsets compatible with our results here, about $45\%$ of their clusters and especially the merging, multiple-clump ones, show larger separations than $10\arcsec$, with a maximum of $\simeq54\arcsec$. This, however, may be a result of either large fractions of unrelaxed clusters in their sample ($\sim60\%$), and, the ensemble of different SL techniques used for the comparison, many of which pre-assume or iterate for the DM centre while adopting a symmetric DM distribution (see references therein), which may be unrealistic given the perturbed and complex matter distribution seen especially in unrelaxed clusters. In that sense, such offsets or even the known discrepancy between mass estimates from lensing and X-ray \citep[e.g.][]{Allen1998L-XrayDisc,Richard2010locuss20}, may not be surprising \citep[see][]{Shan2010OriginOffsets}. This in fact is a crucial point to make here, as we implied above: In our work we do not fit to the data a symmetric DM distribution for which the effective DM centre may be in practice different than the DM peak, whose offset from the BCG we characterise here.

Several studies have also examined the offsets between the BCG and the X-ray peak or centroid \citep[e.g.][]{LinMohr2004,Maughan2008clusterEv}. Recently, \citet{MannEbeling2012} examined the X-ray peak and centroid offsets from the BCG, in 108 of the most luminous X-ray clusters, with the goal of constraining the evolution with redshift of the cluster merger fraction, so that they also characterised the evolution of such offsets, with redshift. Similar to our result, they also found that the distribution is (roughly) log-normal, and centred at 11.5 kpc ($H_{0}=70$ km~s$^{-1}$~Mpc$^{-1}$) for the offset of the BCG from the X-ray peak, overall similar to, or of the same order of, the peak centre we find for our BCG-DM offset distribution: 12.7 $h^{-1}$ kpc. For the BCG offset from the X-ray centroid, their peak is $\sim$twice as large. In addition, they found an evolution for these offsets with redshift, so that higher separations are generally expected for higher redshift clusters, probably as a sign of higher merging fraction, similar to the (tentaive) evolution we observe here between the DM peak and the BCG. This evolution generally agrees also with other complementary studies such as (a different X-ray sample) brightness centroid shifts, metallicity, brightness profile steepness, or other similar relations found in cluster evolution works \citep[e.g.][and references therein]{Maughan2008clusterEv}.

A similarly interesting question which should be investigated quantitatively, is the correlation between the offset magnitude and the degree of relaxation. While in our work the BCG to DM peak offset may imply, generally, the degree of relaxation, other independent relaxation measures would be needed in order to obtain at least a reference sample for one to find a correlation between these. In a similar manner, the degree of relaxation have been defined by various (simulation-based) works in the literature \citep[e.g.][]{Thomas200tauCDM,Neto2007}, where the offset between the potential minimum and the centre of mass, constitutes one of the measures for relaxation \citep[see][]{Lemze2011AnisProfiles}, however these require knowledge of both. It would be therefore worth developing other independent relaxation measures to compare with, in future studies.

In a similar context, and although in our previous analyses we have successfully covered large ranges of these parameters, once many more clusters are analysed in detail using multiple images, in combination also with WL data where possible, it would be interesting to test whether the procedure described here still applies to new limits of richness, luminosity, relaxation, concentration, mass, et cetera.


We also note, that an alternative smoothing process to consider, could be based on a Gaussian filter smoothing rather a fit to a 2D polynomial as performed here, since the latter could suffer from various edge effects and other boundary condition artefacts (although, importantly, with no noticeable effect on the central, DM peak location; the length scale is an order of magnitude different). The reason we did not use a Gaussian smoothing, but the 2D spline interpolation, is that the latter method is well vetted and works remarkably well in reconstructing many mass distributions, as seen in our previous lensing analyses mentioned throughout. We are, however, in the process of examining the Gaussian smoothing alternative, with the goal of establishing the range of relevant Gaussian-filter widths to be implemented (e.g. as priors) in the future analyses, for comparison.

In our work here we made an additional attempt to identify the level of misidentified and grossly miscentred clusters. We found that substantial part of the misidentified clusters show a typical offset behaviour in the $q$-$S$ parameter plane, of which we made use to exclude such clusters from the fit and estimate their portion in the whole sample. Accordingly, we found that about $10\%$ of the clusters in the catalogue may be misidentified, which is \emph{smaller} but of the same general order as found in the previous studies mentioned throughout. In that sense, also, note that we do not attempt to characterise the offset distribution of these misidentified clusters, nor can we distinguish them from real, grossly miscenetred clusters, if these indeed exist and survived our cleaning procedure. Both by the offset log-normal shape we find here which drops for higher offsets, and the difference between the $10\%$ noise level we find and the level found in other studies, the effect of both largely offsetted and misidentified clusters in the remaining sample (after cleaning the misidentified clusters we spotted in our procedure) on our results, is expected to be minimal (see also \S \ref{offsetNoiseS1}).

\section{Summary}\label{disc}

In this work we investigated the distribution of BCG to DM centre offsets, with an automated mass-modelling method (usually used for lensing analyses), based on the well-tested assumption that light traces mass. The modelling includes a lumpy component representing the cluster galaxies, and a smooth component (of the galaxy distribution) representing the DM. The success of this assumption in describing the mass distributions of lensing clusters assures us that the mass model is able to determine the effective DM peak or centre, that can be then compared with the position of the BCG. Unlike typical lensing analyses, we do not work here with multiple images to constrain the fit, and so we take into account a full range of smoothing (polynomial fit) degrees as a flat prior, to cover all options of offsets per cluster.

Few previous studies have dealt with the DM to BCG offset, although these were either performed using numerical simulations mock catalogues probing the high-end of the misidentified or significantly miscentred clusters \citep[][]{Johnston2007offset,HilbertWhite2010} finding a 2D Gaussian width of $\sigma=0.34-0.42~ h^{-1}$ Mpc, or by a WL analysis of 25 clusters \citep[][2D Gaussian width of $\sigma=0.09~ h^{-1}$ Mpc, and a secondary tail with $\sigma=0.42~ h^{-1}$ Mpc]{Oguri2010_25clusters}, which unfortunately entails a relatively low resolution to compare to our results. Recent work by \citet{George2012offsets}, establishes in 129 galaxy groups (although as massive as $<10^{14}~M_{\odot}$) that the typical WL halo centre to BCG offset is smaller, around 75 kpc. In addition, BCG to X-ray (or SZ) peak and centroid offsets have been also characterised \citep[e.g.][]{LinMohr2004,Maughan2008clusterEv,MannEbeling2012,Sehgal2012offsetsSZBCG,Song2012OffsetsSZ}, as well as BCG or X-ray, to SL centre offsets previously derived for small samples of clusters \citep[e.g.][]{Smith2005,Richard2010locuss20,Shan2010OriginOffsets}. Here, we have taken upon to concentrate and characterise the projected offset distribution with a high-resolution analysis, incorporating a mass-modelling method used for detailed SL analyses, and performed over an unprecedentedly large sample of 10,000 SDSS clusters drawn from the \citet{Hao2010GMBCG_cat} catalogue. Such an automated study, in addition to charcterising the offset distribution which is of its own importance, could help identify highly perturbed, potentially merging `bullet-like'' clusters and to statistically constrain the DM cross-section, especially when complementary X-ray observations are in hand \citep[e.g.][]{Clowe2006Bullet,Bradac2008macsbullet,Merten2011}.

The first of the two main results of this work, is therefore the offset distribution (Figures \ref{1dDistLog} and \ref{1dDistLogarcsec}), and the fact it has a non-zero peak. The offsets are well described by a log-normal distribution, with $\langle \log_{10}(\Delta~[h^{-1} $Mpc$])\rangle=-1.895^{+0.003}_{-0.004}$, and $\sigma=0.501\pm0.004$ ($95\%$ confidence levels), or $\langle \log_{10}(\Delta [\arcsec])\rangle=0.564\pm0.005$, and $\sigma=0.475\pm0.007$. Here the mean log offsets correspond to $\Delta \simeq 12.7~h^{-1}$kpc, and $\Delta \simeq 3.6\arcsec$, respectively. Note that these offsets are significantly smaller than those probed by \citet{Johnston2007offset} and \citet{Oguri2010_25clusters}, but more similar to the various BCG to X-ray peak and centroid offsets, or BCG and X-ray to SL centre offsets, mentioned above \citep[see also][]{George2012offsets}. However the centre of DM in the aforementioned WL or SL analyses is usually determined by a fit to certain symmetric model, e.g., \citet[][NFW]{Navarro1996}. It is not surprising therefore that the results will be somewhat different, as the fit of the SL or WL data to an NFW profile, for example, is susceptible to the overall shape of the 2D mass distribution (and out to larger scales in the WL case); while in our smoothing procedure we do not fit to a specific model, and explicitly examine the offset between the BCG and highest peak centre of the smooth DM map. Our results show that the effect on, say, future WL analyses, of the ``real'' DM-BCG offset, is therefore expected to be relatively small (see Eq. 9 and Figure 4 in \citealt{Johnston2007offset}; see also \citealt{George2012offsets} and references therein), although we leave the exact assessment of this effect for relevant future work.

To establish how significant is the typical offset being different than zero, we performed the Student t-test. We examine if the (both linear and log) distributions could have been drawn from a Gaussian distribution with a mean of zero. To do this we first convert the offsets back to one-dimensional Gaussian random number by multiplying with a random phase. The absolute t-statistics is 0.1(0.19), with an error probability of 0.42(0.46), that the linear(logarithmic) distribution is drawn from a Gaussian with a zero mean. We therefore conclude that although relatively small, the typical offset is significantly non-zero. In addition, the effective DM centre we adopt here, namely the peak of the smoothed light distribution representing the DM, can constitute a natural and alternative definition of cluster centers for optically-selected cluster catalogues.

The second main result of this work, is that the offset distribution shows on average, a steady trend with redshift (although given the distribution widths the trend is rendered so far insignificant, see Figures \ref{growthMPC} and \ref{growthArcsec}). Higher redshift clusters show generally larger separations, both in physical scales and angular scales, and more mildly, also smaller distribution widths. This evolution with redshift, although tentative, may be related evidently to cluster evolution and relaxation processes, as less relaxed clusters which are younger and therefore more often found in higher redshifts, have a more spread-out mass distribution and tend correspondingly to exhibit larger BCG-DM offsets. In other words, the matter in lower-redshift clusters has had more time to fall into and form a well-defined potential well. This result is consistent with brightness centroid shifts, metallicity, brightness profile steepness, and other similar relations found in cluster evolution works (e.g. \citealt{Maughan2008clusterEv}, and references therein; \citealt{MannEbeling2012}, see also \citealt{Zitrin2011d}).

The origin of the (slightly) smaller distribution widths in higher-redshift clusters is not perfectly clear, although similarly, it may well be a characteristic for the diversity of the cluster population: higher-redshift clusters will be mostly unrelaxed, whereas lower-redshift clusters can be more of a mixed population in a somewhat wider variety of relaxation states. In addition to our assessment of the DM-BCG offsets distribution, the tentative trends seen here as a function of redshift are real, the results of this work could in turn constitute a complementary and independent measure of cluster evolution history, in future studies.

It should also be mentioned that the results of our work are highly coupled to the cluster catalogue in use and its inherent uncertainties. Future comparisons with independent catalogues, whose clusters are not necessarily optically-selected, should aid in establishing further our results.

\section*{acknowledgments}
We thank the anonymous reviewer of this work for valuable comments. AZ acknowledges support by contract research ``Internationale Spitzenforschung II-1'' of the Baden W\"urttemberg Stiftung, and is grateful for very useful discussions with Yoel Rephaeli, Matthias Redlich, Ole Host and Mauricio Carrasco, and to Jiangang Hao for the GMBCG catalogue and Narciso Ben\'itez for useful conversion templates. We acknowledge use of some publicly-available MATLAB scripts, by Eran Ofek and Salman Rogers. This work was supported in part by the FIRST program "Subaru
Measurements of Images and Redshifts (SuMIRe)", World Premier
International Research Center Initiative (WPI Initiative), MEXT,
Japan, and Grant-in-Aid for Scientific Research from the JSPS
(23740161). Funding for the SDSS and SDSS-II has been provided by the Alfred P. Sloan Foundation,
the Participating Institutions, the National Science Foundation, the U.S. Department of Energy,
the National Aeronautics and Space Administration, the Japanese Monbukagakusho, the Max
Planck Society, and the Higher Education Funding Council for England. The SDSS Web Site is
http://www.sdss.org/.
The SDSS is managed by the Astrophysical Research Consortium for the Participating Institutions.
The Participating Institutions are the American Museum of Natural History, Astrophysical
Institute Potsdam, University of Basel, University of Cambridge, Case Western Reserve University,
University of Chicago, Drexel University, Fermilab, the Institute for Advanced Study, the Japan
Participation Group, Johns Hopkins University, the Joint Institute for Nuclear Astrophysics, the
Kavli Institute for Particle Astrophysics and Cosmology, the Korean Scientist Group, the Chinese
Academy of Sciences (LAMOST), Los Alamos National Laboratory, the Max-Planck-Institute for
Astronomy (MPIA), the Max-Planck-Institute for Astrophysics (MPA), New Mexico State University,
Ohio State University, University of Pittsburgh, University of Portsmouth, Princeton University,
the United States Naval Observatory, and the University of Washington.

\bibliographystyle{mn2eNEW}
\bibliography{outDan2}

\bsp
\label{lastpage}

\end{document}